\documentclass{article}
\usepackage{dcase2021_techrep,amsmath,graphicx,url,times,booktabs, tabularx}

\title{Multi-source transformer architectures \\ for audiovisual scene classification}

\name{Wim Boes, Hugo Van hamme}
\address{
ESAT, KU Leuven \\
wim.boes@esat.kuleuven.be, hugo.vanhamme@esat.kuleuven.be}

\begin{document}

\ninept
\maketitle

\begin{sloppy}

\begin{abstract}
In this technical report, the systems we submitted for subtask 1B of the DCASE 2021 challenge, regarding audiovisual scene classification, are described in detail. They are essentially multi-source transformers employing a combination of auditory and visual features to make predictions. These models are evaluated utilizing the macro-averaged multi-class cross-entropy and accuracy metrics.

In terms of the macro-averaged multi-class cross-entropy, our best model achieved a score of 0.620 on the validation data. This is slightly better than the performance of the baseline system (0.658). 

With regard to the accuracy measure, our best model achieved a score of 77.1\% on the validation data, which is about the same as the performance obtained by the baseline system (77.0\%).
\end{abstract}


\section{Introduction}
\label{sec:intro}

Subtask 1B of DCASE 2021~\cite{challenge1B} is dedicated to audiovisual scene classification. Models are ranked using the macro-averaged multi-class cross-entropy, which is further explained in Section 3.

For this work, we were influenced by the prior use of transformers in the context of environmental event classification with audiovisual data~\cite{audiovisual_transformers}. We also were inspired by the employment of multi-source transformer architectures in the context of machine translation with multiple languages~\cite{multisrc}.

In Section~\ref{sect:models}, the submitted systems are described. In Section~\ref{sect:setup}, we go into the experimental setup. Next, in Section~\ref{sect:results}, we report the obtained results, and finally, we draw a conclusion in Section~\ref{sect:conclusion}.

\section{Models}
\label{sect:models}

In this section, the submitted models are elaborated upon.

\subsection{Architecture}

The architecture of the models is visualized in Figure~\ref{fig:architecture}.

\begin{figure*}[!ht]
    \centering
    \includegraphics[width=\textwidth]{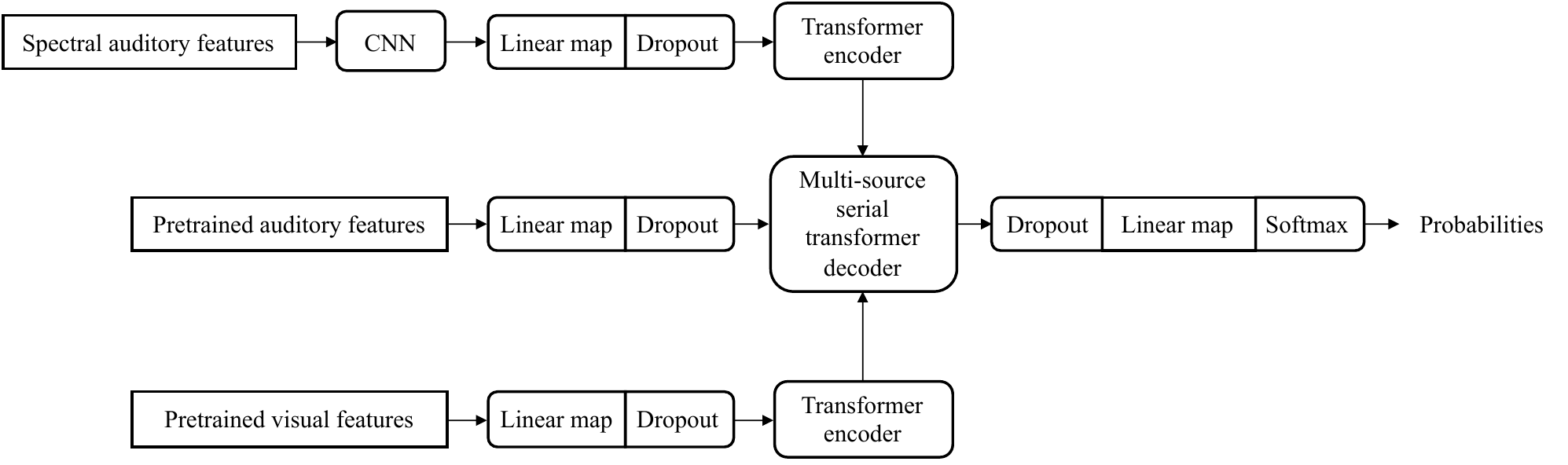}
    \caption{Architecture of submitted models.}
    \label{fig:architecture}
\end{figure*}

The input to the system is a combination of three inputs: spectral auditory features, pretrained auditory features and pretrained visual features. How these are obtained from audiovisual scene recordings is further explained in Section~\ref{sect:setup}.

The spectral auditory features are first processed by a convolutional neural network (CNN), consisting of four blocks. Each block comprises five layers: a convolutional layer, a batch normalization layer~\cite{batchnorm}, a ReLU activation layer, a dropout layer~\cite{dropout} (with a drop rate of 33\%) and an average pooling layer. 

Each convolutional layer uses a square kernel of size 3 and a stride of 1. The amount of output channels are equal to 12, 24, 48 and 96 for the first, second, third and fourth blocks respectively.

\begin{table}[!ht]
\caption{Kernel sizes and strides of pooling layers in CNN}
\label{tab:poolhyper}
\centering
\begin{tabular}{@{}lc@{}}
\toprule
\textbf{Block} & \textbf{Kernel size $=$ stride} \\
\midrule
0 & (3, 4) \\
1-2 & (2, 4) \\
3 & (1, 2) \\
\bottomrule
\end{tabular}
\end{table}

The kernel sizes and strides of the pooling operations are listed in Table~\ref{tab:poolhyper}. The first and second numbers of each tuple relate to the time and frequency axes respectively.

At the end of the last block, the frequency-related dimension of the spectral input has been reduced to one and the corresponding axis can therefore be discarded. 

The sequences of convolutional auditory, pretrained auditory and pretrained visual features are linearly mapped to embeddings of size 96. The outputs of these operations are additionally passed through dropout layers with a drop rate of 33\%.

The embeddings extracted from the convolutional auditory and pretrained visual features are then (separately) passed through a transformer encoder consisting of three layers.

Next, a multi-source serial transformer decoder with three layers is used. This module is a simple extension of the regular transformer decoder allowing for more than two inputs by using multiple multi-head cross-attention blocks (in series) instead of just one. 

In the considered model, the embeddings extracted from the pretrained auditory features constitute the queries of the multi-source serial transformer decoder. The keys and values of the first and second multi-head cross-attention blocks in this decoder come from the transformer encodings based on the convolutional auditory and pretrained visual features respectively.

All transformer components described above use three attention heads, fully connected layers with 96 units and dropout rates of 0.1.

As explained in Section~\ref{sect:setup}, the pretrained auditory features fed as input to the architecture are ``sequences'' of one vector. Thus, the output of the transformer decoder also consists of just one element. 

This vector is passed through a dropout layer, a linear map is computed, and finally, the softmax activation function is applied to obtain the output audiovisual scene probabilities.

\subsection{Loss function}
The loss used to train these models is the categorical cross entropy. 

\subsection{Mean teacher ensemble}

In this work, we employ an ensembling method based on the mean teacher principle~\cite{meanteacher}: The architecture described above is duplicated but is not trained in the regular way. Instead, its parameters are computed as the exponential moving average of the weights of the original structure with a multiplicative decay factor of 0.999 per training iteration. This ``teacher'' model can effectively be regarded as an averaged version of the ``student'' system.

\section{Experimental setup}
\label{sect:setup}

\subsection{Data}
\label{sect:data}

The multi-class data set for problem 1B of DCASE 2021~\cite{baselinedata1B} consists of synchronized auditory and visual scene recordings. There are 10 possible, mutually exclusive audiovisual scenes. 

The training partition consists of 8696 files of 10 seconds, which are cut into 86960 samples of one second. The validation set contains 72000 examples with a length of one second.

\subsection{Preprocessing}

To get spectral maps of the audio segments, we resampled them to 22.050 kHz and performed peak amplitude normalization. Next, log mel spectrograms with 128 bins were computed using a window of 2048 samples and a hop length of 368 samples. Per-frequency bin standardization was done as a final step. For the sound snippets of one second, this resulted in sequences of 600 features.

To get pretrained auditory features, we resampled the audio to 16 kHz and extracted log mel spectrograms with 128 frequency bins utilizing a window of 400 samples and a hop length of 160 samples. The resulting spectral maps were then fed to vggish~\cite{vggish}, a model for video tag classification based on audio, pretrained using a preliminary version of the YouTube-8M data set~\cite{YouTube8M}. The 128-dimensional outputs of the last fully connected layer of this network were used as pretrained auditory features. For an audio recording of one second, this resulted in a ``sequence'' of one vector.

To get pretrained visual features, the frames of the visual files (recorded at 29.97 fps) were passed through VGG16~\cite{VGG16}, a network for image classification, pretrained on the ImageNet data set~\cite{ImageNet}. The 4096-dimensional outputs of the last fully connected layer of this model were used as pretrained visual features. For a video of one second, this resulted in a sequence of 30 features.

\subsection{Data augmentation}

In order to diminish the risk of overfitting, we employed data augmentation during the training of the considered models. More specifically, we used mixup~\cite{mixup} with an activation probability of 50\%. The mixing ratios were randomly sampled from a beta distribution with shape parameters set to 0.2.

\subsection{Training and evaluation}

All of the models were trained and evaluated using PyTorch~\cite{pytorch}.

\subsubsection{Training hyperparameters}

All systems were trained for 30 epochs with a batch size of 128. Every sample in each batch originated from a different training file.

Adam~\cite{adam} was employed to train the weights of the models. Learning rates were ramped up exponentially from 0 to 0.00025 for the first 675 optimization iterations. Subsequently, they decayed multiplicatively at a rate of 0.999 per training step.

\subsubsection{Evaluation metrics}

For this challenge, two evaluation metrics are used: macro-averaged multi-class (categorical) cross-entropy and accuracy. 

After each training epoch, these metrics were calculated using the probabilities produced by the teacher models. The best scores were retained and are reported in Section~\ref{sect:results}.

\section{Results}
\label{sect:results}

The baseline model for this particular challenge~\cite{baselinedata1B} achieved a macro-averaged multi-class cross-entropy measure of 0.658 and an accuracy of 77.0\% on the validation data.

Our best model achieved a cross-entropy of 0.620 on the validation data, which is a slight improvement over the baseline. 

With regard to the accuracy measure, our best model achieved a score of 77.1\% on the validation data, which is about the same as the performance obtained by the baseline system.

\section{Conclusion}
\label{sect:conclusion}

In this technical report, we described the systems we submitted for subtask 1B of the DCASE 2021 challenge and reported their results. 

The best macro-averaged cross-entropy and accuracy scores achieved by our models were equal to 0.620 and 77.1\% respectively.

\section{Acknowledgment}
\label{sec:ack}

This work is supported by a PhD Fellowship of Research Foundation Flanders (FWO-Vlaanderen).

\bibliographystyle{IEEEtran}
\bibliography{refs}

\begin{thebibliography}{10}
\providecommand{\url}[1]{#1}
\def\UrlFont{\rmfamily}
\providecommand{\newblock}{\relax}
\providecommand{\bibinfo}[2]{#2}
\providecommand\BIBentrySTDinterwordspacing{\spaceskip=0pt\relax}
\providecommand\BIBentryALTinterwordstretchfactor{4}
\providecommand\BIBentryALTinterwordspacing{\spaceskip=\fontdimen2\font plus
\BIBentryALTinterwordstretchfactor\fontdimen3\font minus
  \fontdimen4\font\relax}
\providecommand\BIBforeignlanguage[2]{{%
\expandafter\ifx\csname l@#1\endcsname\relax
\typeout{** WARNING: IEEEtran.bst: No hyphenation pattern has been}%
\typeout{** loaded for the language `#1'. Using the pattern for}%
\typeout{** the default language instead.}%
\else
\language=\csname l@#1\endcsname
\fi
#2}}

\bibitem{challenge1B}
S.~Wang, T.~Heittola, A.~Mesaros, and T.~Virtanen, ``{Audio-visual scene
  classification: analysis of DCASE 2021 Challenge submissions},'' \emph{arXiv
  preprint arXiv:2105.13675}, 2021.

\bibitem{audiovisual_transformers}
W.~Boes and H.~Van~hamme, ``{Audiovisual transformer architectures for
  large-scale classification and synchronization of weakly labeled audio
  events},'' in \emph{{Proceedings of the 27th ACM International Conference on
  Multimedia}}, 2019, pp. 1961--1969.

\bibitem{multisrc}
J.~Libovick{\`y}, J.~Helcl, and D.~Mare{\v{c}}ek, ``{Input Combination
  Strategies for Multi-Source Transformer Decoder},'' in \emph{{Proceedings of
  the Third Conference on Machine Translation: Research Papers}}, 2018, pp.
  253--260.

\bibitem{batchnorm}
S.~Ioffe and C.~Szegedy, ``{Batch Normalization: Accelerating Deep Network
  Training by Reducing Internal Covariate Shift},'' in \emph{Proceedings of
  ICML}, 2015, pp. 448--456.

\bibitem{dropout}
N.~Srivastava, G.~Hinton, A.~Krizhevsky, I.~Sutskever, and R.~Salakhutdinov,
  ``{Dropout: A Simple Way to Prevent Neural Networks from Overfitting},''
  \emph{The Journal of Machine Learning Research}, vol.~15, no.~1, pp.
  1929--1958, 2014.

\bibitem{meanteacher}
A.~Tarvainen and H.~Valpola, ``{Mean teachers are better role models:
  Weight-averaged consistency targets improve semi-supervised deep learning
  results},'' in \emph{Advances in Neural Information Processing Systems},
  2017, pp. 1195--1204.

\bibitem{baselinedata1B}
S.~Wang, A.~Mesaros, T.~Heittola, and T.~Virtanen, ``{A Curated Dataset of
  Urban Scenes for Audio-Visual Scene Analysis},'' in \emph{{2021 IEEE
  International Conference on Acoustics, Speech and Signal Processing
  (ICASSP)}}, 2021.

\bibitem{vggish}
S.~Hershey, S.~Chaudhuri, D.~P. Ellis, J.~F. Gemmeke, A.~Jansen, R.~C. Moore,
  M.~Plakal, D.~Platt, R.~A. Saurous, B.~Seybold, \emph{et~al.}, ``{CNN
  Architectures for Large-Scale Audio Classification},'' in \emph{2017 IEEE
  International Conference on Acoustics, Speech and Signal processing
  (ICASSP)}, 2017, pp. 131--135.

\bibitem{YouTube8M}
S.~Abu-El-Haija, N.~Kothari, J.~Lee, P.~Natsev, G.~Toderici, B.~Varadarajan,
  and S.~Vijayanarasimhan, ``{Youtube-8M: A Large-Scale Video Classification
  Benchmark},'' \emph{arXiv preprint arXiv:1609.08675}, 2016.

\bibitem{VGG16}
K.~Simonyan and A.~Zisserman, ``{Very Deep Convolutional Networks for
  Large-Scale Image Recognition},'' in \emph{International Conference on
  Learning Representations}, 2015.

\bibitem{ImageNet}
J.~Deng, W.~Dong, R.~Socher, L.-J. Li, K.~Li, and L.~Fei-Fei, ``{ImageNet: A
  Large-Scale Hierarchical Image Database},'' in \emph{CVPR09}, 2009.

\bibitem{mixup}
H.~Zhang, M.~Cisse, Y.~N. Dauphin, and D.~Lopez-Paz, ``{mixup: Beyond empirical
  risk minimization},'' \emph{arXiv preprint arXiv:1710.09412}, 2017.

\bibitem{pytorch}
A.~Paszke, S.~Gross, F.~Massa, A.~Lerer, J.~Bradbury, G.~Chanan, T.~Killeen,
  Z.~Lin, N.~Gimelshein, L.~Antiga, A.~Desmaison, A.~Kopf, E.~Yang, Z.~DeVito,
  M.~Raison, A.~Tejani, S.~Chilamkurthy, B.~Steiner, L.~Fang, J.~Bai, and
  S.~Chintala, ``{PyTorch: An Imperative Style, High-Performance Deep Learning
  Library},'' in \emph{Advances in Neural Information Processing Systems},
  2019, pp. 8024--8035.

\bibitem{adam}
D.~P. Kingma and J.~Ba, ``{Adam: A Method for Stochastic Optimization},''
  \emph{arXiv preprint arXiv:1412.6980}, 2014.

\end{thebibliography}

\end{sloppy}
\end{document}